\documentclass[submission, Phys]{SciPost}

\hypersetup{
    colorlinks,
    linkcolor={red!50!black},
    citecolor={blue!50!black},
    urlcolor={blue!80!black}
}

\usepackage{graphicx}
\usepackage{amsmath,amssymb}
\usepackage{bm}
\usepackage{braket}

\usepackage[bitstream-charter]{mathdesign}
\DeclareSymbolFont{usualmathcal}{OMS}{cmsy}{m}{n}
\DeclareSymbolFontAlphabet{\mathcal}{usualmathcal}

\newcommand{\pdiff}[2]{\frac{\partial #1}{\partial #2}}

\newcommand{\new}{\nonumber\\}
\newcommand{\bx}{\bm{x}}
\newcommand{\bq}{\bm{q}}

\newcommand{\tu}{\tilde{u}}

\newcommand{\txi}{\tilde{\xi}}
\newcommand{\abs}[1]{\left|#1\right|}
\newcommand{\ave}[1]{\left\langle #1\right\rangle}

\begin{document}
\begin{center}{\Large \textbf{Harmonic chain far from equilibrium: single-file diffusion, long-range
order, and hyperuniformity\\ }}
\end{center}

\begin{center}
Harukuni Ikeda\textsuperscript{1$\star$}
\end{center}

\begin{center}
{\bf 1} Department of Physics, Gakushuin University, 1-5-1 Mejiro, Toshima-ku, Tokyo 171-8588, Japan
\\
${}^\star$ {\small \sf harukuni.ikeda@gakushuin.ac.jp}
\end{center}

\begin{center}
\today
\end{center}


\section*{Abstract}
{\bf
In one dimension, particles can not bypass each other. As a consequence,
the mean-squared displacement (MSD) in equilibrium shows sub-diffusion
${\rm MSD}(t)\sim t^{1/2}$, instead of normal diffusion ${\rm
MSD}(t)\sim t$. This phenomenon is the so-called single-file
diffusion. Here, we investigate how the above equilibrium behaviors are
modified far from equilibrium. In particular, we want to uncover what
kind of non-equilibrium driving force can suppress diffusion and achieve
the long-range crystalline order in one dimension, which is prohibited
by the Mermin-Wagner theorem in equilibrium. For that purpose, we
investigate the harmonic chain driven by the following four types of
driving forces that do not satisfy the detailed balance: (i) temporally
correlated noise with the noise spectrum $D(\omega)\sim
\omega^{-2\theta}$, (ii) conserving noise, (iii) periodic
driving force, and (iv) periodic deformations of particles. For the
driving force (i) with $\theta>-1/4$, we observe ${\rm MSD}(t)\sim
t^{1/2+2\theta}$ for large $t$. On the other hand, for the driving
forces (i) with $\theta<-1/4$ and (ii)-(iv), MSD remains finite. As a
consequence, the harmonic chain exhibits the crystalline order even in
one dimension.
Furthermore, the density fluctuations of the model are highly suppressed
in a large scale in the crystal phase.  This phenomenon is known as
hyperuniformity.  We discuss that hyperuniformity of the noise
fluctuations themselves is the relevant mechanism to stabilize the
long-range crystalline order in one dimension and yield hyperuniformity
of the density fluctuations.}

\vspace{10pt}
\noindent\rule{\textwidth}{1pt}
\tableofcontents\thispagestyle{fancy}
\noindent\rule{\textwidth}{1pt}
\vspace{10pt}

\section{Introduction}
In one-dimensional many-particle systems, the particles cannot bypass
one another if the interactions are strong enough. As a consequence,
the mean-squared displacement (MSD) in equilibrium shows sub-diffusion
${\rm MSD}(t)\sim
t^{1/2}$~\cite{alexander1978,karger1992,kollmann2003,lin2005,taloni2017single,tirthankar2022},
instead of normal diffusion ${\rm MSD}(t)\sim
t$~\cite{zwanzig2001nonequilibrium}. This phenomenon is known as
single-file diffusion. The simplest model to observe single-file
diffusion is the one-dimensional harmonic chain, where point-like
particles on a line are connected by harmonic
springs~\cite{alexander1978}. The harmonic chain is often recognized as
a toy model of a one-dimensional
crystal~\cite{ashcroft2022solid,altland2010condensed}. However, as
proved by Mermin and Wagner, the long-range order cannot exist in one
and two dimensions in equilibrium if interactions are
short-ranged~\cite{mermin1968,mermin1969,altland2010condensed}. This
implies that the particles will diffuse away from their lattice
positions after a sufficiently long time. As a consequence, MSD of the
harmonic chain grows as ${\rm MSD}\sim t^{1/2}$, as in the case of
standard single-file diffusion~\cite{alexander1978}. The aim of this
manuscript is to discuss how the above equilibrium behaviors are changed
if the model is driven by athermal fluctuations violating the detailed
balance. In particular, we show that for specific types of athermal
fluctuations, the diffusion is strongly suppressed, and as a
consequence, the harmonic chain can have the long-range crystalline
order even in one dimension.

Our model also provides an ideal playground to investigate
hyperuniformity far from equilibrium. Hyperuniformity is a phenomenon
that the large-scale fluctuations of physical quantities are anomalously
suppressed.  In particular, hyperuniformity of the density fluctuations
is characterized by the vanishing of the static structure factor $S(q)$
in the limit of the small wave number $q$: $\lim_{q\to
0}S(q)=0$~\cite{torquato2018hyperuniform}. Hyperuniformity has been
observed in perfect crystals at zero temperature~\cite{kim2018},
quasicrystals~\cite{oguz2017,oǧuz2019hyperuniformity}, ground states of
quantum systems~\cite{feynman1954,reatto1967,Torquato2008}, periodically
driven emulsions~\cite{weijs2015emergent}, chiral active
matter~\cite{lei2019hydrodynamics,huang2021circular,hyperchiral2022,kuroda2023microscopic},
and so on~\cite{hexner2015,hexner2017noise}. Interestingly, a recent
numerical study reported that hyperuniformity of out-of-equilibrium
systems can also suppress the critical fluctuations and stabilize the
crystaline order even in two dimension~\cite{leonardo2023}, which is
prohibited by the Mermin-Wagner theorem in
equilibrium~\cite{mermin1968,mermin1969}.
So far, most of the theoretical studies of hyperuniformity far from
equilibrium have been conducted on low densities much below the crystal
phase~\cite{lei2019hydrodynamics,Lei2019,kuroda2023microscopic}.  We
believe that the harmonic chain plays the role of the minimal model for
investigating how hyperuniformity appears and stabilizes the crystalline
order in low-dimensional systems far from equilibrium.

The manuscript is organized as follows. In Sec.~\ref{214728_15Sep23}, we
introduce the model and define a few important physical quantities.
Then, we perform case studies for the following four types of driving
forces that do not satisfy the detailed balance.

Firstly, in Sec.~\ref{214807_15Sep23}, we consider the temporally
correlated noise with the power-law noise spectrum $D_q(\omega)\sim
\omega^{-2\theta}$. Although the model may appear somewhat artificial,
it allows us to systematically investigate how the temporal correlations
enhance or suppress the diffusion and yield long-range crystalline order
and hyperuniformity by continuously changing the value of $\theta$.  We
show that for $\theta>-1/4$, the mean-squared displacement behaves as
${\rm MSD}\sim t^{1/2+2\theta}$ for large $t$.  For $\theta<-1/4$, on
the contrary, the diffusion is completely suppressed, and MSD converges
to a finite value in the long time limit. As a consequence, the model
exhibits the long-range crystalline order even in one dimension, which
is prohibited in equilibrium~\cite{mermin1968,mermin1969}.  Furthermore,
we show that the static structure factor $S(q)$ for a small wave number
$q$ behaves as $S(q)\sim q^{-4\theta}$ in the crystal phase.
In the crystal phase, $S(q)\to 0$ in the limit $q\to 0$, meaning that
the density fluctuations show
hyperuniformity~\cite{torquato2018hyperuniform}. We discuss that
hyperuniformity of the density fluctuations is a consequence of temporal
hyperuniformity of the noise, {\it i.e.}, the fluctuations of the noise
vanish in the long-time scale $\lim_{\omega\to 0}D_q(\omega)=0$.

Secondly, in Sec.~\ref{214904_15Sep23}, we consider the systems driven
by conserving noise to investigate the effects of the spatial
correlation of the noise. In previous work, Hexner and Levine have shown
that for a system driven by conserving noise, the density fluctuations
are highly suppressed and exhibit
hyperuniformity~\cite{hexner2017noise}, due to hyperuniformity of the
noise itself~\cite{ikeda2023out}. Recently, Galliano {\it et
al.}~\cite{leonardo2023} have shown that the suppression of the
fluctuations also stabilizes the long-range crystalline order even in
two dimension, which is prohibited by the Mermin-Wagner theorem in
equilibrium. Does the crystalline order also emerge in one dimension?
We show that the harmonic chain driven by the conserving noise indeed
possesses the crystalline order~\cite{leonardo2023}. We also show that
the static structure factor for small $q$ behaves as $S(q)\sim q^2$,
meaning that the density fluctuations show
hyperuniformity~\cite{torquato2018hyperuniform}, as observed in previous
works~\cite{hexner2017noise,leonardo2023}. We discuss that
hyperuniformity of the density fluctuations is a consequence of spatial
hyperuniformity of the noise itself, {\it i.e.}, $\lim_{q\to
0}D_q(\omega)=0$.

Thirdly, in Sec.~\ref{214520_3Sep23}, we investigate a periodically
driven system. For that purpose, we consider chiral active particles
confined in a narrow one-dimensional channel and connected with harmonic
springs~\cite{Callegari2019,liebchen2022chiral}. We show that MSD
oscillates with the same frequency as that of the driving force, and the
crystalline order parameter takes a finite value. We also show $S(q)\sim
q^2$ for $q\ll 1$, meaning that the model shows hyperuniformity, as
previously observed in chiral active matter in two
dimension~\cite{lei2019hydrodynamics,huang2021circular,hyperchiral2022,kuroda2023microscopic}.
We discuss that hyperuniformity of the density fluctuations is a
consequence of temporal hyperuniformity of the noise itself, {\it i.e.},
$\lim_{\omega\to 0}D_q(\omega)=0$.

Finally, in Sec.\ref{224807_15Sep23}, we consider periodically deforming
particles in one dimension, which was originally introduced as a model
to describe dense biological tissues~\cite{tjhung2017}. The driving
force of the model oscillates with the constant frequency and
simultaneously has the conserving nature. The Fourier spectrum of
the driving force satisfies $\lim_{q\to 0}D_q(\omega)=0$ and
$\lim_{\omega\to 0}D_q(\omega)=0$, meaning that the driving force is
spatio-temporally hyperuniform. Under the harmonic approximation, the
model can be reduced to the one-dimensional harmonic chain with the
oscillating natural lengths. We show that MSD oscillates with the same
frequency as that of the driving force, and the crystalline order
parameter takes a finite value. We also show that the model exhibits
stronger hyperuniformity than those of the conserving
noise and periodic driving forces: $S(q)\sim q^4$ for $q\ll
1$~\cite{torquato2018hyperuniform}.

The above four case studies demonstrate that the temporal and/or spatial
hyperuniformity of the driving force yield hyperuniformity of the
density fluctuations and can stabilize the crystalline order even in one
dimension~\cite{ikeda2023out,ikeda2023does}, while the positive
correlation enhances the diffusion. In Sec.~\ref{225058_15Sep23}, we
summarize those results and give more quantitative discussion for the
connection between the strength of hyperuniformity and existence of the
crystalline order in low-dimensional systems.

\section{Model and physical quantities}
\label{214728_15Sep23} Here, we introduce the model and define a few
important physical quantities.

\subsection{Model}
We consider the harmonic chain driven by the following
dynamics~\cite{zwanzig2001nonequilibrium}:
\begin{align}
\dot{x}_j = K(x_{j+1}+x_{j-1}-2x_j) + \xi_j(t),\
j = 1,\cdots, N,
 \label{105141_29Aug23}
\end{align}
where $\{x_j\}_{j=1,\cdots, N}$, $\{\xi_j\}_{j=1,\cdots, N}$, $K$, and
$N$ denote the positions of the particles, driving forces, spring
constant, and number of particles, respectively. We impose the periodic
boundary condition $x_{N+1}=x_1$. We investigate the model in the
center-of-mass frame, which is, in practice, equivalent to replacing the
noise in (\ref{105141_29Aug23}) as $\xi_j\to \xi_j-\sum_{k=1}^N
\xi_k/N$.  Let $R_j$ be the equilibrium position of the $j$-th particle. The
dynamical equation for the displacement $u_j=x_j-R_j$ is then written as
\begin{align}
\dot{u}_j = K(u_{j+1}+u_{j-1}-2u_j) + \xi_j(t).
    \label{dyn}
\end{align}
It is convenient to introduce the Fourier and inverse Fourier
transformations of the displacement
$u_j(t)$~\cite{zwanzig2001nonequilibrium,ashcroft2022solid}:
\begin{align}
&u_j(t) = \frac{1}{\sqrt{N}}\sum_{q}\tu_{q}(t)e^{iq R_j},
&\tu_q(t) = \frac{1}{\sqrt{N}}\sum_{j=1}^{N}u_j(t)e^{-iqR_j},
\end{align}
where $q\in \left\{\frac{2\pi k}{Na} \right\}_{k=-N/2,\cdots,
N/2-1}$ if $N$ is even,
and $q\in \left\{\frac{2\pi k}{Na} \right\}_{k=-(N-1)/2,\cdots,
(N-1)/2}$ if $N$ is odd. Eq.~(\ref{dyn}) is diagonalized in the Fourier space:
\begin{align}
\dot{\tu}_q(t) = -\lambda_q \tu_q(t) + \txi_q(t),
\label{qk}
\end{align}
where 
\begin{align}
\lambda_q = 2K\left[1-\cos(aq)\right],
\end{align}
and 
\begin{align}
\txi_q(t) = \frac{1}{\sqrt{N}}\sum_{j=1}^{N}\xi_j(t)e^{-iqR_j}.
\end{align}
Note that $\tilde{\xi}_{q=0}(t)=0$ in the center-of-mass frame. We
assume that the mean and variance of $\txi_k(t)$ are given by
\begin{align}
&\ave{\txi_q(t)} = 0,
&\ave{\txi_q(t)\txi_{q'}(t')} = \delta_{q,-q'}D_q(t-t').
\end{align}
We initialize the system with $x_j=ja$ at time $t=t_0$. Since the center
of mass is fixed, the equilibrium position is given by $R_j=ja$, if the
crystalline state is stable. Then, Eq.~(\ref{qk}) can be solved directly
to give
\begin{align}
\tilde{u}_q(t) = \tilde{u}_q(t_0)e^{-\lambda_q(t-t_0)}
 + \int_{t_0}^t e^{-\lambda_q(t-s)}
\tilde{\xi}_q(s)ds.
\end{align}
We take the limit $t_0\to -\infty$ so that the system reaches the steady
state at $t=0$. The two point correlation in the Fourier space is then
calculated as
\begin{align}
\ave{\tu_q(\omega)\tu_{q'}(\omega')}  =
2\pi\delta_{q,-q'}\delta(\omega+\omega')\frac{D_q(\omega)}{\omega^2 + \lambda_q^2}.
\end{align}
The inverse Fourier transform w.r.t. $\omega$ yields
\begin{align}
\ave{\tu_q(t)\tu_{-q}(0)}  = \frac{1}{2\pi}\int_{-\infty}^{\infty} d\omega
e^{i\omega t}\frac{D_q(\omega)}{\omega^2 + \lambda_q^2}
= \frac{1}{\pi}\int_0^{\infty} d\omega \frac{D_q(\omega)\cos(\omega t)}{\omega^2 + \lambda_q^2},
\label{131429_31Aug23}
\end{align}
where we used the time-reversal symmetry of the correlation: $D_q(\omega)=D_q(-\omega)$.

\subsection{Mean-squared displacement}

Using the Parseval's identity, the mean-squared displacement in the
thermodynamic limit $N\to\infty$ is calculated as follows:
\begin{align}
{\rm MSD}(t)&=\frac{1}{N}\sum_{j=1}^N \ave{\left(u_j(t)-u_j(0)\right)^2}\new 
&=
\frac{1}{N}\sum_{q}\ave{\left(\tu_{q}(t)-\tu_{q}(0)\right)
 \left(\tu_{-q}(t)-\tu_{-q}(0)\right)}\new 
&= \frac{2}{N}\sum_{q}\frac{1}{\pi}\int_0^\infty d\omega D_{q}(\omega)
 \frac{1-\cos(\omega t)}{\omega^2+\lambda_{q}^2}\new
&=\frac{1}{\pi^2}\int_{-\pi/a}^{\pi/a}adq \int_0^{\infty}d\omega 
D_q(\omega) \frac{1-\cos(\omega t)}{\omega^2 +\lambda_q^2},
\label{020259_3Sep23}
\end{align}
where we have replaced the summation for $q$ with an integral for $q\in
(-\pi/a,\pi/a)$.

\subsection{Order parameter}
To quantify the crystalline order, we observe the Fourier component of
the density at the reciprocal wave number $q=2\pi/a$~\cite{mermin1968}:
\begin{align}
O = \frac{1}{N}\ave{\sum_{j=1}^N e^{i\frac{2\pi}{a}x_j}}    
= \ave{e^{i\frac{2\pi u_1}{a}}}
= \ave{\cos\left(\frac{2\pi u_1}{a}\right)}.
\end{align}
The order parameter vanishes $O=0$ for disordered liquid-like
configurations, while $O>0$ for crystals~\cite{mermin1968}. In
equilibrium, the distribution of $u_1$ becomes a Gaussian. Therefore,
the order parameter is calculated as
$O=\exp\left[-\frac{2\pi^2\ave{u_1^2}}{a^2}\right]$.  However at finite
temperature, the fluctuation diverges $\ave{u_1^2}\to\infty$ in the
thermodynamic limit $N\to\infty$, leading to $O\to
0$~\cite{nishimori2010elements}. Therefore, the thermal fluctuations
destroy the crystalline order in one dimension, which is consistent with
the Mermin-Wagner theorem~\cite{mermin1968,mermin1969}.  Of course, 
the theorem does not hold in systems far from equilibrium. Indeed, in the
later sections, we will show several examples of the out-of equilibrium
driving forces that preserve the crystalline order even in one
dimension.

\subsection{Hyperuniformity}
For perfect crystals at zero temperature, the static structure factor
$S(q)$ in the limit of the small wave number vanishes: $\lim_{q\to
0}S(q)=0$. In other words, the density fluctuations are highly
suppressed for small $q$. This property is referred to as
hyperuniformity~\cite{torquato2018hyperuniform}. In equilibrium, on the
contrary, $\lim_{q\to 0}S(q)$ converges to a finite value, namely, the
thermal fluctuations destroy hyperuniformity~\cite{kim2018}. Does
hyperuniformity survive under the athermal fluctuations considered in
this work? To answer this question, we calculate $S(q)$ for $q\ll
1$ as follows~\cite{kim2018}:
\begin{align}
S(q) &= \ave{\frac{1}{N}\abs{\sum_{j=1}^N e^{iq x_j}}^2}\new 
&\approx \frac{1}{N}\ave{\abs{\sum_{j=1}^Ne^{iqR_j}}^2}
+ q^2 \ave{\frac{1}{N}\abs{\sum_{j=1}^N u_je^{iqR_j}}^2}\new 
&= S_0(q) + q^2 \ave{\tu_q \tu_{-q}}\new
&\approx  q^2 \ave{\tu_q \tu_{-q}}.\label{170347_4Sep23}
\end{align}
where $S_0(q)=\ave{\abs{\sum_{j=1}^Ne^{iqR_j}}^2}/N$ denotes the static
structure factor of the one-dimensional lattice, which has delta peaks
at $q=2\pi n/a$, $n=0,1,2,\cdots$ and can be ignored for sufficiently
small but finite $q$. Eq.~(\ref{170347_4Sep23}) allows us to discuss
hyperuniformity from the scaling of $\ave{\tu_q \tu_{-q}}$ for small
$q$. Note that the expansion (\ref{170347_4Sep23}) is verified only in
the crystal phase. In the fluid phase, even for small $q$, $u_j$ can be
very large so the expansion by $qu_j$ breaks.

The interaction potential of the harmonic chain is diagonalized in the
Fourier space: $V_N = \sum_{q}\frac{\lambda_q }{2}\tu_{q} \tu_{-q}$.
From the law of equipartition, one obtains
$\ave{\tu_{q}\tu_{-q}}=T/\lambda_q$ in equilibrium at temperature
$T$. This leads to $S(q)\approx T/(a^2K)$ for $q\ll 1$, meaning that the
thermal fluctuations destroy hyperuniformity~\footnote{The classical harmonic
chain in equilibrium does not have the crystalline order in one
dimension at finite temperature~\cite{mermin1968}. However a careful
treatment can justify Eq.~(\ref{170347_4Sep23}) even in the fluid phase,
see Ref.~\cite{kim2018}. The same is true for quantum harmonic chain at
zero temperature~\cite{reatto1967}.}. On the contrary, hyperuniformity
is often observed in systems driven by athermal fluctuations, where the
law of equipartition does not hold. The simplest and well-known example
is the quantum harmonic chain:
$H=\sum_q\frac{\tilde{p}_q\tilde{p}_{-q}}{2}+\sum_q
\frac{\lambda_q}{2}\tu_{q} \tu_{-q}$, where the momentum $\tilde{p}_q$
satisfies the canonical commutation relation
$[\tu_{q},\tilde{p}_{q'}]=i\delta_{q,-q'}\hbar$~\cite{feynman1954,reatto1967}. On
the ground state, the distribution of $\tu_{q}$ is a Gaussian of zero
mean and variance
$\ave{\tu_{q}\tu_{-q}}=\hbar/(2\sqrt{\lambda_q})$~\cite{greiner2012thermodynamics}.
Therefore, the static structure factor is approximated as $S(q)\approx
q\hbar /(2a\sqrt{K})$ for $q\ll 1$~\cite{feynman1954,reatto1967}: the
model exhibits hyperuniformity $\lim_{q\to 0}S(q)=0$. In the subsequent
sections, we will show that hyperuniformity can also emerge in classical
systems far from equilibrium.

\section{Temporally correlated noise}
\label{214807_15Sep23}

For the first example of the athermal fluctuations, we here consider the
temporally correlated noise. The model allows us to understand how the
time correlations of the noise yield
hyperuniformity of the density fluctuations, and how these properties
affect the diffusion and long-range order.

\subsection{Settings}
Here we consider the Gaussian color noise of zero mean and variance
\begin{align}
\ave{\xi_i(t)\xi_j(t')} = \delta_{ij}D(t).
\end{align}
In previous work, single-file diffusion of active particles has been
investigated~\cite{dolai2020universal}.  In that case, the correlation
of the noise $D(t)$ decays exponentially, which leads to the same
scaling as that in equilibrium ${\rm MSD}(t)\sim t^{1/2}$ for $t\gg
1$~\cite{dolai2020universal}. As we will see below, the scaling is
altered if $D(t)$ has the power-law tail. We assume that 
the noise spectrum is written as
\begin{align}
&D(\omega) = 2T\abs{\omega}^{-2\theta}\cos(\theta \pi),
&-1/2 < \theta < 1/2.\label{111300_6Sep23}
\end{align}
Here, the pre-factor $\cos(\theta\pi)$ has been chosen to simplify the
final result, and $\theta$ is restricted to $-1/2<\theta<1/2$ to
converge the correlation function, as we will see later. When
$\theta=0$, the model satisfies the detailed
balance~\cite{zwanzig2001nonequilibrium}, and thus $\xi_i$ can be
identified with the thermal noise at temperature $T$. For $\theta>0$,
the noise has the positive correlation that decays algebraically for
large $t$: $D(t)\sim 1/t^{1-2\theta}$. On the contrary, for $\theta<0$,
$\lim_{\omega\to 0}D(\omega)=0$, meaning that the noise fluctuations are
highly suppressed in the long-time scale. Namely, the noise is
temporally hyperuniform.

The power-law spectrum Eq.~(\ref{111300_6Sep23}) of the noise appears
for non-equilibrium systems showing self-organized
criticality~\cite{per1987,Newman2005} and is often referred to as the
$1/f$ noise~\cite{milotti2002}. The power-law spectrum also appears for
the Fourier spectrum of quasi-periodic patterns. In
Ref.~\cite{oǧuz2019hyperuniformity}, the authors showed that the Fourier
spectrum of one-dimensional quasi-periodic patterns exhibits the
power-law behavior for small $\omega$ with $\theta \in [-3/2,1]$. Also,
in Ref.~\cite{kim2018}, the authors argued that small perturbations to
one-dimensional periodic patterns yield the power-law spectrum for
$\theta\in [-1,0]$. Therefore, the model driven by the noise with the
correlation Eq.~(\ref{111300_6Sep23}) would give useful insights for
quasi-periodically and periodically driven systems.

The power-law spectrum (\ref{111300_6Sep23}) has been often used in the
context of anomalous diffusion~\cite{eliazar2009unified}.  A single
free particle driven by the noise, $\dot{x}=\xi(t)$, exhibits ${\rm
MSD}\propto t^{1+2\theta}$ for large time $t$.

\subsection{Mean-squared displacement}
\begin{figure}[t]
\begin{center}
\includegraphics[width=10cm]{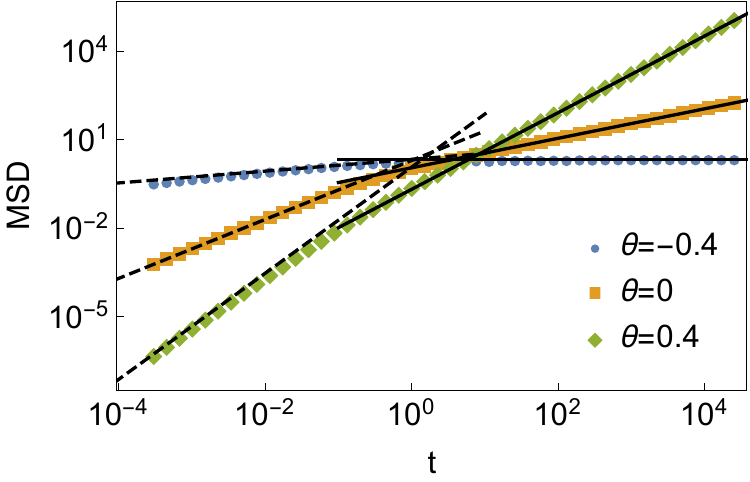} \caption{Mean-squared
displacement of harmonic chain driven by temporally correlated
noise. Markers denote exact results. Dashed and solid lines represent
short and long-time asymptotic behaviors, respectively. For simplicity,
we set $K=1$ and $T=1$.}
				       \label{224630_29Aug23}
\end{center}
\end{figure}

In the thermodynamic limit $N\to\infty$, the mean-squared displacement
Eq.~(\ref{020259_3Sep23}) is calculated as
\begin{align}
{\rm MSD}(t)
&=\frac{2T}{\pi^2\sec({\pi\theta})}\int_{-\pi/a}^{\pi/a}adq \int_0^{\infty}d\omega 
\abs{\omega}^{-2\theta}\frac{1-\cos(\omega t)}{\omega^2 + \left[2K(1-\cos(aq))\right]^2}. 
\end{align}
 The integral w.r.t $q$ can be performed as
\begin{align}
 \int_{-\pi/a}^{\pi/a}\frac{adq }{\omega^2 + \left[2K(1-\cos(aq))\right]^2}
= 2\pi \sqrt{\frac{\omega+\sqrt{\omega^2+16K^2}}{2\omega^3(\omega^2+16K^2)}}
\sim 
 \begin{cases}
 \frac{\pi}{\sqrt{2K\omega^3}} & \abs{\omega}\ll 1 \\
 \frac{2\pi}{\omega^2} & \abs{\omega}\gg 1.
 \end{cases} 
\end{align}
Using this result, 
one can deduce the scaling of MSD for $t\ll 1$ as follows:
\begin{align}
{\rm MSD}(t) \sim At^{1+2\theta}\ (t\ll 1),
\end{align}
where $A$ denotes a constant.
This scaling agrees with that of a free-particle driven by the temporally
correlated noise~\cite{eliazar2009unified}.
For $t\gg 1$ and $\theta>-1/4$, we get 
\begin{align}
{\rm MSD}(t) \sim
Bt^{\frac{1}{2}+2\theta}\ (t\gg 1),\label{020958_27Sep23}
\end{align}
where $B$ denotes a constant. For $\theta=0$, one recovers the scaling
of single-file diffusion in equilibrium ${\rm MSD}\sim
t^{1/2}$~\cite{alexander1978,taloni2017single}. For $\theta<-1/4$, ${\rm
MSD}$ in the long time limit converges to a finite value:
$\lim_{t\to\infty}{\rm MSD}(t) = 2\ave{u_1^2}$. We plot ${\rm MSD}$ for
several $\theta$ in Fig.~\ref{224630_29Aug23}.

Eq.~(\ref{020958_27Sep23}) can be understood from a simple
scaling argument. To see this, we consider the continuum limit of
Eq.~(\ref{dyn}):
\begin{align}
\dot{u}(x,t) = K\nabla^2 u(x,t) + \xi(x,t),\label{020532_27Sep23},
\end{align}
where the noise correlation is given by $\ave{\xi(x,t)\xi(x',t')} =
\delta(x-x')D(t-t')$. To analyze the model in the large spatio-temporal
scale, we consider the following scaling transformations: $x\to bx$,
$t\to b^{z_t}t$, $u\to b^{z_u}u$. Assuming that all terms in
Eq.~(\ref{020532_27Sep23}) have the same scaling dimension, we obtain
$z_t=2$ and $z_u=1/2+2\theta$~\cite{burger1989,ikeda2023out}. This leads
to ${\rm MSD}\sim u^2 \sim b^{2z_u}\sim t^{2z_u/z_t}\sim
t^{1/2+2\theta}$, which is consistent with Eq.~(\ref{020958_27Sep23}).

\subsection{Order parameter}
\begin{figure}[t]
\begin{center}
\includegraphics[width=15cm]{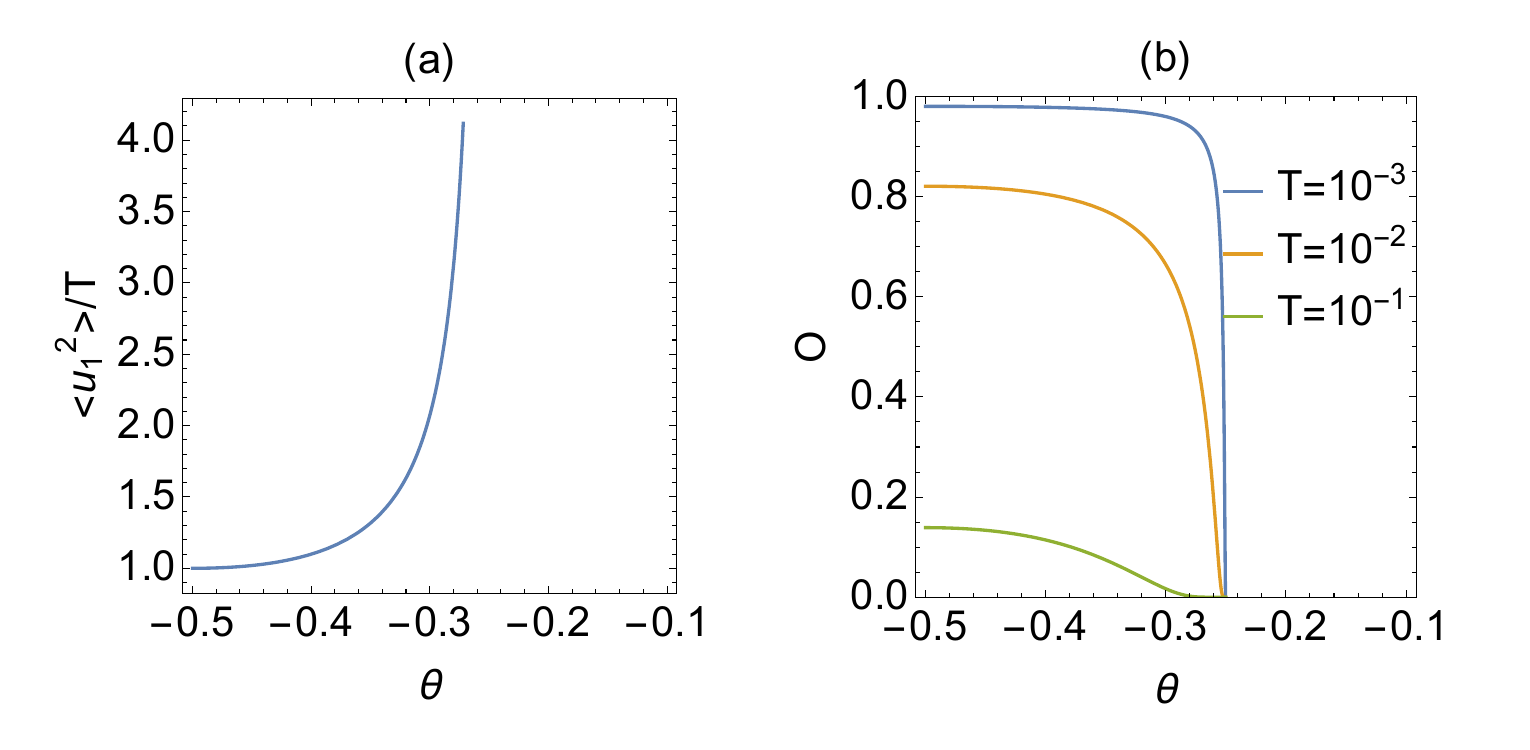} \caption{Physical quantities of
harmonic chain driven by temporally correlated noise.  (a) $\theta$
dependence of the fluctuation $\ave{u_1^2}$. $\ave{u_1^2}$ has a finite
value for $\theta<-1/4$ and diverges at $\theta=-1/4$. (b) $\theta$
dependence of the order parameter $O$ for several temperatures. For
$\theta<-1/4$, $O>0$, while for $\theta\geq -1/4$, $O=0$. For
simplicity, we here set $K=1$ and $a=1$.}  \label{160256_28Aug23}
\end{center}
\end{figure}

The equal time correlation in the Fourier space is 
\begin{align}
\ave{\tu_{q} \tu_{-q}}
 = \frac{1}{\pi}\int_0^{\infty}\frac{D(\omega)d\omega}{\omega^2+\lambda_q^2}
= \frac{2T}{\pi\sec(\theta\pi) (\lambda_q)^{1+2\theta}}\int_{0}^{\infty}dx
\frac{\abs{x}^{-2\theta}}{x^2+1}
 = \frac{T}{(\lambda_q)^{1+2\theta}}.\label{123947_6Sep23}
\end{align}
Note that the integral converges only when $-1/2<\theta<1/2$. When
$\theta=0$, we recover the law of equipartition:
\begin{align}
 \ave{\tu_{q} \tu_{-q}} = \frac{T}{\lambda_q}.
\end{align}
In real space, we get 
\begin{align}
\ave{u_1^2} = \frac{1}{N}\sum_{j=1}^N \ave{u_j^2}
 = \frac{1}{N}\sum_{q} \ave{\tu_{q} \tu_{-q}}
 = \frac{T}{N}\sum_{q} \frac{1}{\lambda_q^{1+2\theta}}
= \frac{T}{\pi}
 \int_{-\pi/a}^{\pi/a} adq \frac{1}{\left[2K(1-\cos(aq))\right]^{1+2\theta}}.
\end{align}
For $\theta<-1/4$, the integral converges to
\begin{align}
\frac{\ave{u_1^2}}{T} =
 -\frac{\theta\Gamma(-1/2-2\theta)}
 {2^{1+4\theta}K^{1+2\theta}\pi^{1/2}\Gamma(1-2\theta)},
\end{align}
while for $\theta\geq -1/4$, $\ave{u_1^2}\to\infty$, see
Fig.~\ref{160256_28Aug23}~(a). Since $\xi_i$ is a Gaussian random
number, the solution of the linear differential equation
Eq.~(\ref{dyn}), $u_1$, also becomes a Gaussian random
number~\cite{zwanzig2001nonequilibrium}. Therefore, the order parameter
can be calculated as
\begin{align}
O = 
\ave{e^{i\frac{2\pi u_i}{a}}} = \exp\left[-\frac{2\pi^2}{a^2}\ave{u_1^2}\right].
\end{align}
We plot $O$ in Fig.~\ref{160256_28Aug23}~(b). The order parameter $O$
has a finite value for $\theta<-1/4$, meaning that the model has the
long-range crystalline order even in one dimension. For $\theta>-1/4$, $O=0$, implying that the
diffusion destroys the crystalline order.

\subsection{Hyperuniformity}

In the crystal phase ($\theta<-1/4$), $S(q)$ for $q\ll 1$ is
approximated as
\begin{align}
S(q)\approx q^2 \ave{\tu_{q} \tu_{-q}}=T\frac{q^2}{\lambda_q^{1+2\theta}}
\approx \frac{Tq^{-4\theta}}{(Ka^2)^{1+2\theta}}.\label{124701_6Sep23}
\end{align}
In the crystal phase, $\lim_{q\to 0}S(q)=0$, meaning that the system is
hyperuniform~\cite{torquato2018hyperuniform}. For $\theta<0$, the noise
spectrum satisfies $\lim_{\omega\to 0}D(\omega)=0$, meaning that the
noise is temporally hyperuniform.  The above result implies that
temporal hyperniformiy of the noise leads to hyperuniformiy of the
density fluctuations in the crystal phase.

The above analysis is limited for $\theta>-1/2$ because the correlation
Eq.~(\ref{123947_6Sep23}) diverges for $\theta\leq -1/2$. This
ultraviolet divergence can be removed by introducing a phenomenological
cut-off to the spectrum.  For that purpose, we consider the modified
power-spectrum:
\begin{align}
D(\omega) =
\begin{cases}
 C\abs{\omega}^{-2\theta} &\abs{\omega }< \omega_c\\ 
 0 & {\rm otherwise}
\end{cases},
\label{md}
\end{align}
where $C$ denotes a constant. The
correlation Eq.~(\ref{123947_6Sep23}) for $q\ll 1$ can be
calculated as
\begin{align}
 &\ave{\tu_{q} \tu_{-q}} =
\frac{T}{\pi}\int_0^{\omega_c}d\omega\frac{C\abs{\omega}^{-2\theta}}{\omega^2+\lambda_q^2}
\sim \frac{T}{\pi}\int_0^{\omega_c}d\omega\abs{\omega}^{-2\theta-2}, &\theta<-1/2,\label{134951_6Sep23} 
\end{align}
where $\omega_c$ denotes the cut-off frequency. The integral converges
to a constant value for $\theta<-1/2$. Therefore, we get $S(q)\approx
q^2 \ave{\tu_{q}\tu_{-q}}\sim q^2$, which is consistent with the limit
$\theta\to -1/2$ of Eq.~(\ref{124701_6Sep23}).

One can also investigate the effects of the power-law spatial
correlation $D_q(\omega)\sim q^{-2\rho}$. Since the analysis is very
parallel to that in this section, we just shortly summarize the main
consequences of the power-law spatial correlation in
Sec.~\ref{225058_15Sep23}. From the next sections, we shall focus on
more concrete examples.

\section{Conserving noise}
\label{214904_15Sep23}

In the previous section, we have observed that temporal hyperuniformity of
the noise leads to hyperuniformity of the density fluctuations and also
stabilizes the crystalline order even in one dimension. In this section,
we shall show that spatial hyperuniformity of the noise can also yield
hyperuniformity of the density fluctuations and stabilize the
crystalline order~\cite{hexner2017noise}.

\subsection{Settings}

Hyperuniformity is a phenomenon that the fluctuations of physical
quantities become much smaller than what would be expected from the
central limit theorem. Hyperuniformity has been reported in various
systems, such as crystals,
quasicrystals~\cite{kim2018,oǧuz2019hyperuniformity,chen2021stone}, and
chiral active
matter~\cite{huang2021circular,hyperchiral2022,kuroda2023microscopic}. In
general, the physical mechanisms causing hyperuniformity can differ
depending on the details of the systems. However, Hexner and Levine have
pointed out that hyperuniformity can universally appear for
out-of-equilibrium systems driven by conserving
noise~\cite{hexner2017noise}.
Here, we argue that the same scenario also holds in a one-dimensional
system driven by the conserving noise.

The conserving noise in the continuum limit $\xi(x,t)$ is written as
$\xi(x,t)=\partial_x \eta(x,t)$, where $\eta(x,t)$ denotes another white
noise. A simple implementation of this condition is
\begin{align}
\xi_j(t) = \eta_{j}(t)-\eta_{j-1}(t),
\end{align}
where $\eta_j(t)$ is a Gaussian random number of zero mean
and variance:
\begin{align}
\ave{\eta_i(t)\eta_j(t')} = T\delta_{ij}\delta(t-t').
\end{align}
The Fourier component of $\xi_j(t)$ satisfies
\begin{align}
&\ave{\txi_q(t)} = 0,\new 
&\ave{\txi_q(t)\txi_{q'}(t)} = 4\delta_{q,-q'}T\left[1-\cos(aq)\right]\delta(t-t')
=\delta_{q,-q'}D_q(t-t'),
\end{align}
where
\begin{align}
D_q(t) = \frac{2T\lambda_q}{K}\delta(t).\label{051024_17Sep23}
\end{align}
For $q\ll 1$, $\lambda_q\sim q^2$ and thus $D_q(t)\sim q^2$, meaning
that the large-scale spatial fluctuations of the noise are highly
suppressed. In other words, the noise is spatially hyperuniform.

\subsection{Mean-squared displacement}
\begin{figure}[t]
\begin{center}
\includegraphics[width=10cm]{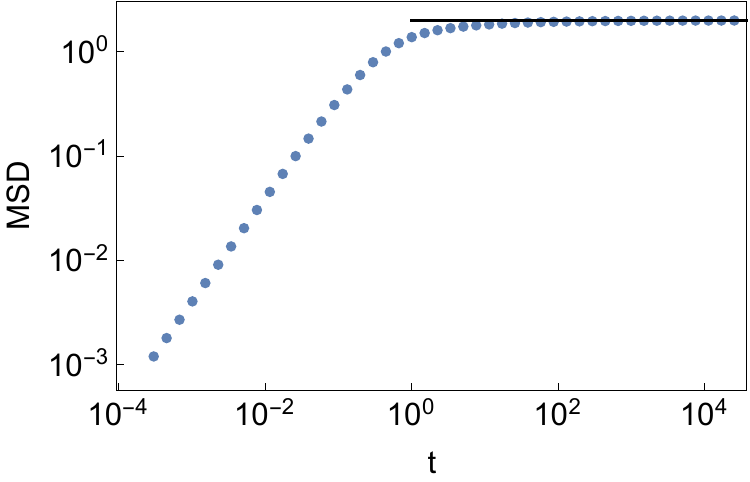} \caption{Mean-squared
displacement of harmonic chain driven by conserving
noise. Markers denote exact results.  Solid lines represent long time
asymptotic behavior: ${\rm MSD}\sim 2\ave{u_1^2}$. For simplicity, 
we set $K=1$ and $T=1$.}  \label{033750_3Sep23}
\end{center}
\end{figure}

In the thermodynamics limit $N\to\infty$, the mean-squared displacement
is calculated as
\begin{align}
{\rm MSD}(t)
&=\frac{2T}{\pi^2K} \int_0^{\infty}d\omega
(1-\cos(\omega t))
\int_{-\pi/a}^{\pi/a}adq
\frac{2K(1-\cos(aq))}{\omega^2 + \left[2K(1-\cos(aq))\right]^2}. 
\end{align}
We plot MSD in Fig.~\ref{033750_3Sep23}. MSD converges to a finite value
in the long-time limit, $\lim_{t\to\infty}{\rm MSD} =
2\ave{u_1^2}$. This means that the particles are localized around their
lattice positions, and thus the model is expected to have the
crystalline order. Below, we confirm that this intuition is correct.

\subsection{Order parameter}
Repeating the same analysis as in Eq.~(\ref{131429_31Aug23}), 
we get 
\begin{align}
\ave{\tu_{q}\tu_{-q}} =
 \frac{1}{\pi}\int_0^{\infty}d\omega \frac{D_q(\omega)}{\omega^2+\lambda_q^2}
 = \frac{2T}{K\pi}\int_0^{\infty}d\omega \frac{\lambda_q}{\omega^2+\lambda_q^2}
=\frac{T}{K}.\label{155446_30Sep23}
\end{align}
The squared deviation from the lattice position is then calculated as
\begin{align}
 \ave{u_1^2} =
 \frac{1}{N}\sum_{q}\ave{\tu_{q}\tu_{-q}} = 
\frac{T}{K}.
\end{align}
Since $\xi_j(t)$ is a Gaussian random number and the model only has the
linear interactions, $u_1$ also follows the Gaussian distribution.  Thus,
the order parameter is
\begin{align}
 O = \frac{1}{N}\ave{\sum_{j=1}^N e^{\frac{2\pi i}{a}x_j}} = \ave{e^{\frac{2\pi i}{a}u_1}}
= \exp\left[-\frac{2\pi^2T}{a^2K}\right].
\end{align}
The order parameter has a finite value, meaning that the model driven by
the conserving noise has the crystalline order even in
one dimension.

\subsection{Hyperuniformity}
Hexner and Levine argued that the density fluctuations are anomalously
suppressed in systems driven by conserving noise~\cite{hexner2017noise}.
To see this, we calculate $S(q)$
for small $q\ll 1$:
\begin{align}
 S(q)\approx q^2 \ave{\tu_{q}\tu_{-q}} = \frac{T}{K}q^2.\label{013507_1Sep23}
\end{align}
$S(q)$ vanishes in the limit $q\to 0$, meaning that the large-scale
density fluctuations are highly suppressed. This is the signature of 
hyperuniformity~\cite{torquato2018hyperuniform}.

Overall, the above results imply that spatial hyperuniformity of the
noise $\lim_{q\to 0}D_q(\omega)=0$ yields hyperuniformity of the density
fluctuations and stabilizes the long-range crystalline order even in
one-dimension.

\subsection{Mapping to Einstein model}
Interestingly, the current model can be mapped into an equilibrium
model. This can be seen by rewriting Eq.~(\ref{qk}) as follows:
\begin{align}
&\pdiff{\tu_{q}(t)}{t} = -\Gamma_q \pdiff{V_{\rm eff}}{\tu_{-q}(t)} + \txi_q(t),
&\ave{\tu_{q}(t)\tu_{q'}(t')} = 2\delta_{q,-q'}T\Gamma_q\delta(t-t'),\label{201701_7Oct23}
\end{align}
where $\Gamma_q=\lambda_q/K$ and $V_{\rm
eff}=\sum_{i=1}^N\frac{K}{2}u_{i}^2$. Eq.~(\ref{201701_7Oct23}) is the
equilibrium Langevin equation satisfying the detailed balance with the
friction coefficient $\Gamma_q$~\cite{zwanzig2001nonequilibrium}. Then,
the steady state distribution follows the Boltzmann distribution:
\begin{align}
&P(u_1,\cdots, u_N)
=\frac{e^{-\frac{V_{\rm eff}}{T}}}{\int \prod_{i=1}^N du_i
 e^{-\frac{V_{\rm eff}}{T}}}.
\end{align}
This is nothing but the Einstein model consisting of $N$ independent
harmonic oscillators of the same frequency $\omega=\sqrt{K}$. The
Einstein model is known to exhibit hyperuniformity~\cite{kim2018}, which
is consistent with Eq.~(\ref{013507_1Sep23}).

\section{Periodically driven system}
\label{214520_3Sep23}

In Sec.~\ref{214807_15Sep23}, we have investigated the effects of
temporal hyperuniformity of the driving force $\lim_{\omega\to
0}D(\omega)=0$. For the extreme case of temporal hyperuniformity, here
we study a periodically driven system, where the Fourier spectrum of the
driving force is strictly zero, $D(\omega)=0$, for $\omega<\omega_0$.

\subsection{Settings}
Here, we consider the periodic driving force. For a concrete example, we
consider chiral active particles in one dimension. Chiral active
particles are particles that exhibit circular
motions~\cite{liebchen2022chiral}. A popular mathematical model to
describe this motion is~\cite{Callegari2019}
\begin{align}
&\dot{x} = \sqrt{2T}\cos\phi + \xi_x,\new
&\dot{y} = \sqrt{2T}\sin\phi + \xi_y,\new
&\dot{\phi} = \omega_0 + \xi_{\phi},
\end{align}
where $\xi_{x,y,\phi}$ denotes the noise.  We are particularly
interested in the limit $\xi_{x,y,\phi}\to 0$, where a chiral active
particle undergoes a purely periodic motion. If the particle is confined
in a one-dimensional channel along the $x$ direction, one can only
consider the motion along that direction:
$\dot{x}=\sqrt{2T}\cos(\omega_0 t+\phi(0))$. How does this periodic
nature of the driving force affect the collective motion? To model the
collective excitation of chiral active particles in one dimension, we
consider the harmonic chain Eq.~(\ref{105141_29Aug23}) driven by the
following periodic function~\cite{ikeda2023does}:
\begin{align}
\xi_j(t) = \sqrt{2T}\cos(\omega_0 t + \psi_j),\label{152007_31Aug23}
\end{align}
where $\psi_j$ denotes a random number uniformly distributed in $[0,2\pi]$.
The mean and variance of $\xi_j(t)$ are then given by 
\begin{align}
&\ave{\xi_j(t)} = 0,\new
&\ave{\xi_i(t)\xi_j(t')} = \delta_{ij}TD(t-t'),\label{002420_22Sep23}
\end{align}
where $D(t)=\cos(\omega_0 t)$. The noise spectrum 
$D(\omega)=\pi\delta(\abs{\omega}-\omega_0)$ vanishes in the
limit of the small frequency: $\lim_{\omega\to 0}D(\omega)=0$. Thus the
noise is temporally hyperuniform. When $\omega_0=0$,
$\xi_j(t)=\sqrt{2T}\cos\psi_j$ plays the role of the random field and
destroys the long-range order in $d\leq 4$ as predicted by Imry and
Ma~\cite{imry1975}. What will happen when $\omega_0\neq 0$?

\subsection{Mean-squared displacement}
\begin{figure}[t]
\begin{center}
\includegraphics[width=10cm]{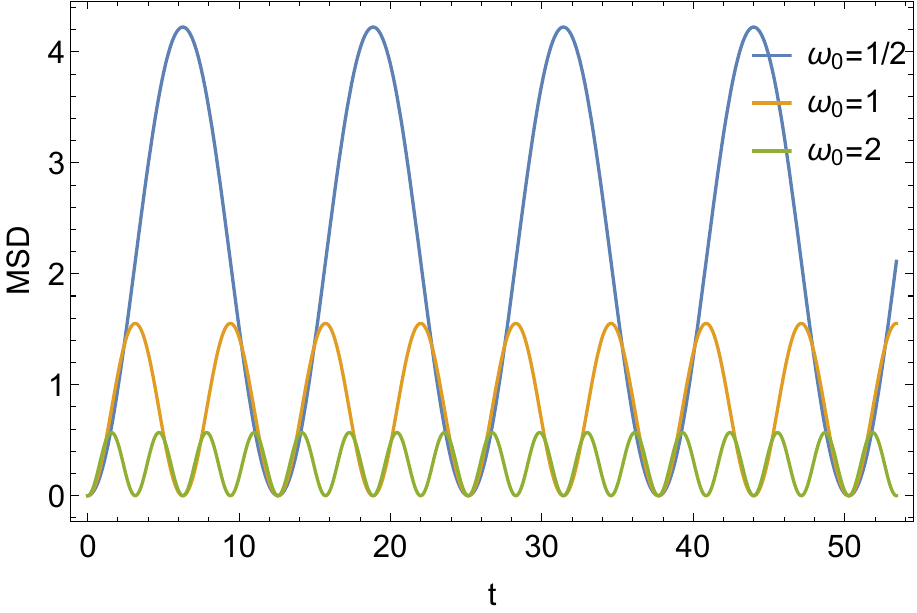}
 \caption{Mean-squared
 displacement of the periodically driven harmonic chain.}
 \label{172742_31Aug23}
\end{center}
\end{figure}

By using Eq.~(\ref{020259_3Sep23}), {\rm MSD} is calculated as
\begin{align}
{\rm MSD}(t) = 2\left[1-\cos(\omega_0 t)\right]\ave{u_1^2}.
\end{align}
MSD oscillates with the frequency of the driving force
$\omega_0$, see Fig.~\ref{172742_31Aug23}.
\begin{figure}[t]
\begin{center}
\includegraphics[width=10cm]{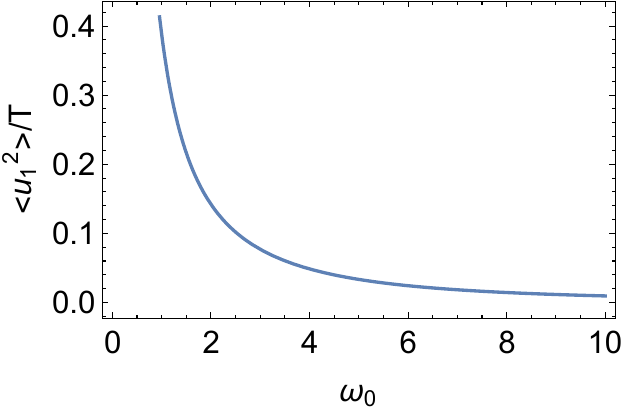} \caption{$\ave{u_1^2}$ of the
priodically driven harmonic chain. $\ave{u_1^2}$ has a finite value for
$\omega_0\neq0$ and diverges in the limit $\omega_0\to 0$.}
\label{151853_31Aug23}
\end{center}
\end{figure}
The fluctuation around the lattice position $\ave{u_1^2}$ is calculated
as
\begin{align}
\ave{u_1^2} =
\frac{T}{\pi}\int_0^{2\pi/a} adq
 \frac{1}{\omega_0^2 + \left[2K(1-\cos(aq))\right]^2},\label{202248_3Sep23}
\end{align}
which has a finite value for $\omega_0\neq 0$ and diverges at
$\omega_0=0$, see Fig.~\ref{151853_31Aug23}. Therefore, the model is
expected to have the crystalline order for $\omega_0\neq 0$.

\subsection{Order parameter}

Because the current driving force Eq.~(\ref{152007_31Aug23}) is not a
Gaussian random variable, one can not easily calculate
$O$. Nevertheless, one can prove the existence of the crystalline order
by using the following inequality~\footnote{To prove the inequality
$\cos(x)\geq 1-x^2/2$, it is convenient to introduce an auxiliary
function $f(x)=\cos(x)-(1-x^2/2)$.  Since $f(x)$ is an even function, it
is sufficient to show $f(x)\geq 0$ for $x\geq 0$, which follows from
$f(0)=0$ and $f'(x)=-\sin(x)+x\geq 0$ for $x\geq 0$.}:
\begin{align}
O =
\ave{e^{i\frac{2\pi u_1}{a}}}=
\ave{\cos\left(\frac{2\pi u_1}{a}\right)} \geq 1- \frac{2\pi^2}{a^2}\ave{u_1^2}.\label{221512_31Aug23}
\end{align}
Since dynamics Eq.~(\ref{dyn}) does not depend on $a$, whether or not
the crystalline order exists is also independent of $a$. So, we chose
$a$ so that $a>\sqrt{2\pi^2\ave{u_1^2}}$. Then
Eq.~(\ref{221512_31Aug23}) leads to $R>0$, meaning that the model has
the crystalline order even in one dimension.

\subsection{Hyperuniformity}

Chiral active particles are known to exhibit hyperuniformity in two
dimension~\cite{lei2019hydrodynamics,huang2021circular,hyperchiral2022,kuroda2023microscopic}.
Does one-dimensional system also exhibit hyperuniformity?  For $q\ll 1$, the static structure factor is calculated as
\begin{align}
S(q)\sim q^2 \ave{\tu_{q}\tu_{-q}} = \frac{Tq^2}{\omega_0^2+\lambda_q^2}\sim 
\begin{cases}
Tq^2/\omega_0^2 & \omega_0>0,\\
Tq^{-2}/(Ka^2)^2 & \omega_0 = 0.
\end{cases}
\label{013554_1Sep23} 
\end{align}
For $\omega_0\neq 0$, the model indeed exhibits hyperuniformity
$S(q)\sim q^2$, as in chiral active matter in two
dimension~\cite{lei2019hydrodynamics,huang2021circular,hyperchiral2022,kuroda2023microscopic}.
The result is also consistent with the temporally correlated noise with
$\theta\leq -1/2$, see Eq.~(\ref{134951_6Sep23}).  This is a reasonable
result because the modified power-law spectrum Eq.~(\ref{md}) converges
to $\lim_{\theta\to-\infty}D(\omega)\propto \delta(\omega-\omega_c)$ in
the limit $\theta\to -\infty$, which agrees with the Fourier-spectrum of
the driving force Eq.~(\ref{002420_22Sep23}). For $\omega_0=0$, on the
contrary, one observes $S(q)\sim q^{-2}$. Therefore, $S(q)$ diverges in
the limit of the small $q$. This anomalous enhancement of the
large-scale density fluctuations is referred to as giant number
fluctuations~\cite{narayan2007long,Zhang2010}. A similar power-law
divergence of $S(q)$ has been previously reported for active matter in
quenched random potentials~\cite{ro2021}.

\section{Periodically deforming particles}
\label{224807_15Sep23}

What will happen if the driving force is a periodic function 
and simultaneously conservative? To answer this
question, we here consider the model introduced by Tjhung and
Berthier~\cite{tjhung2017}.
\subsection{Settings}
\begin{figure}[t]
\begin{center}
\includegraphics[width=10cm]{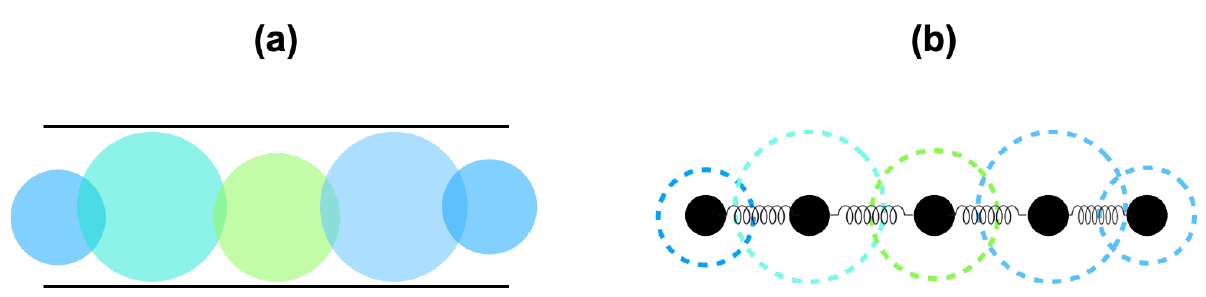} \caption{Schematic figures of
(a) periodically deforming particles in one dimension and (b) harmonic
chain where particle interactions are replaced by linear springs.}
\label{214129_3Sep23}
\end{center}
\end{figure}

 Tissues are often fluidized by periodic
deformations of cells~\cite{Zehnder2015}. To model this behavior, Tjhung
and Berthier introduced periodically deforming
particles~\cite{tjhung2017}. The one-dimensional version of the model is
written as
\begin{align}
&\dot{x}_j(t) = -\pdiff{V_N}{x_j},
&V_N = \sum_{i<j}^Nv(h_{ij}),\label{224643_31Aug23}
\end{align}
where $v(h_{ij})$ denotes the one-sided harmonic potential~\cite{ohern2003}:
\begin{align}
&v(h_{ij}) = \frac{Kh_{ij}^2\Theta(-h_{ij})}{2},
&h_{ij}  = \abs{x_i-x_j}-\frac{r_i(t)+r_j(t)}{2}.
\end{align}
Here the diameter of the $i$-th particle $r_i(t)$ oscillates with
the frequency $\omega_0$~\cite{tjhung2017}:
\begin{align}
r_i(t) = a + \sigma\cos(\omega_0t + \psi_i),
\end{align}
where $\psi_i$ is a random number distributed uniformly in $[0,2\pi]$.
When $\omega_0=0$, $\sigma\cos\psi_i$ plays the role of the
polydispersity, and thus, the model can not have the crystalline
order~\footnote{ For $\omega_0=0$, the driving force
Eq.~(\ref{020427_1Sep23}) becomes a quenched randomness of zero mean and
variance $\ave{\txi_q\txi_{q'}}=T\delta_{q,-q'}\sin(aq)^2$. For $q\ll
1$, $\ave{\txi_q\txi_{q'}}\propto q^2\delta_{q,-q'}$. The Imry-Ma
argument~\cite{imry1975} for the correlated disorder predicts that this
type of disorder prohibits the continuous symmetry breaking for $d\leq
2$~\cite{ikeda2023out}. Therefore, the polydispersity would destroy the
crystalline order in one and two dimensions even without thermal
fluctuations.}. The force term in Eq.~(\ref{224643_31Aug23}) satisfies
Newton's third law~\cite{goldstein2002classical}, which naturally leads
to the conserving driving force as we will see below~\cite{hexner2017noise}.

For sufficiently high density and small $\sigma$,
the harmonic approximation would be justified, and thus the one-sided
harmonic potential would be replaced by the harmonic potential (see Fig.~\ref{214129_3Sep23}):
\begin{align}
v(r_{ij}) \approx \frac{Kh_{ij}^2}{2}.
\end{align}
Taking only the nearest neighbor interactions, one can approximate
Eq.~(\ref{224643_31Aug23}) as
\begin{align}
\dot{x}_j \approx K (x_{j+1}+x_{j-1}-2x_j)
+K\frac{r_{j+1}-r_{j-1}}{2}.
\end{align}
Then, the equation of motion of the displacement $u_j$ is
\begin{align}
\dot{u}_j =K (u_{j+1}+u_{j-1}-2u_j) + \xi_j,\label{013015_1Sep23}
\end{align}
where 
\begin{align}
\xi_j(t) = \frac{K\sigma}{2}\left[\cos(\omega_0t + \psi_{j+1})
-\cos(\omega_0t + \psi_{j-1})\right].\label{020427_1Sep23}
\end{align}
The mean and variance of $\tilde{\xi}_j$ are 
\begin{align}
&\ave{\txi_q(t)} = 0,\new
&\ave{\txi_q(t)\txi_{q'}(t')}  
= T\delta_{q,-q'}D_q(t),
\end{align}
where $T=(K\sigma)^2/4$ and 
\begin{align}
D_q(t)= \sin(aq)^2\cos(\omega_0 t).
\end{align}
The noise spectrum
$D_q(\omega)=\pi\sin(aq)^2\delta(\abs{\omega}-\omega_0)$ vanishes in the
limits of small $\omega$ and/or $q$. Therefore, the noise is
spatio-temporally hyperuniform.

\subsection{Mean-squared displacement}

\begin{figure}[t]
\begin{center}
\includegraphics[width=10cm]{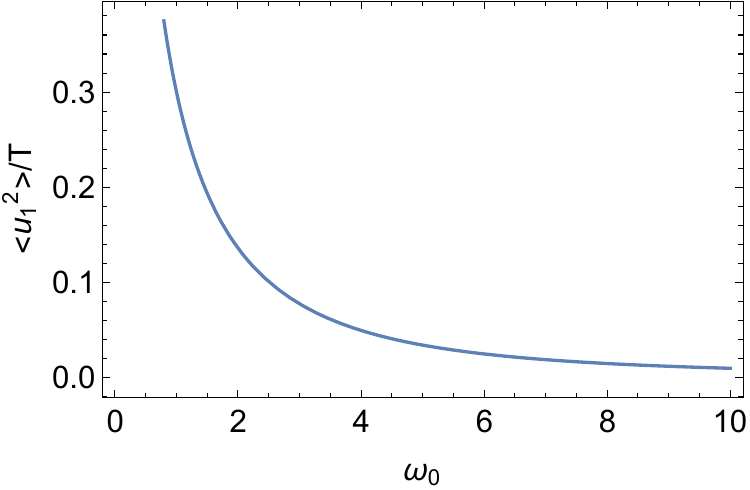} \caption{$\ave{u_1^2}$ of
periodically deforming particles. $\ave{u_1^2}$ has a finite value for
$\omega_0>0$ and diverges in the limit $\omega_0\to 0$.}  \label{011842_1Sep23}
\end{center}
\end{figure}

Repeating the same analysis as in the previous sections, we get
\begin{align}
{\rm MSD}(t) = 2(1-\cos(\omega_0 t))\ave{u_1^2},\label{123820_4Oct23} 
\end{align}
where
\begin{align}
\ave{u_1^2} = \frac{T}{\pi}\int_0^{\pi/a} adq
\frac{(\sin(aq))^2}{\omega_0^2+ \left[2K(1-\cos(aq))\right]^2}.
\end{align}
Eq.~(\ref{123820_4Oct23}) impleis that ${\rm MSD}$ shows the periodic
motion as in the case of the model considered in
Sec.~\ref{214520_3Sep23}. A similar periodic motion of ${\rm MSD}$ has
been previously reported by a numerical simulation of the periodically
deforming particles in two dimension~\cite{Tjhung2017excitation}.

\subsection{Order parameter}
We plot $\ave{u_1^2}$ in Fig.~\ref{011842_1Sep23}. The cage size
$\ave{u_1^2}$ has a finite value for $\omega_0>0$. In this case, using
Eq.~(\ref{221512_31Aug23}) and repeating the same argument as in the
previous section, we can conclude that the model possesses the
crystalline order. In the limit $\omega_0\to 0$, the cage size diverges
$\ave{u_1^2}\to\infty$, and thus one can not prove the existence of the
crystalline order. This is a natural result because when $\omega_0=0$,
the polydispersity $\sigma\cos\psi_i$ destroys the crystalline order.

\subsection{Hyperuniformity}
For small $q\ll 1$,
the static
structure factor  is
\begin{align}
 S(q)\sim q^2 \ave{\tu_{q} \tu_{-q}} = \frac{Tq^2\sin(aq)^2}{\omega_0^2+\lambda_q^2}
 \sim
 \begin{cases}
  Ta^2q^4/\omega_0^2 & \omega_0 > 0\\
  T/(Ka)^2  & \omega_0 = 0.
 \end{cases}\label{033634_17Sep23}
\end{align}
For $\omega_0>0$, we observe $S(q)\sim q^4$, which is much smaller than
the result of the conserving noise Eq.~(\ref{013507_1Sep23}) and
periodic driving force Eq.~(\ref{013554_1Sep23}). This is a consequence
of the fact that the driving force Eq.~(\ref{020427_1Sep23}) is a
periodic function and simultaneously conservative. For $\omega_0=0$,
$S(q)$ converges to a finite value in the limit $q\to 0$, meaning that
the polydispersity destroys hyperuniformity.

Note, we used the fact that $S_0(q)=0$ to derive
Eq.~(\ref{033634_17Sep23}), see Eq.~(\ref{170347_4Sep23}). However, this
condition is not satisfied for amorphous
solids~\cite{aikeda2015}, and polydisperse systems studied in previous
works~\cite{tjhung2017,Tjhung2017excitation}. Our theory predicts that
hyperuniformity is observed only in crystal phases of monodisperse
systems.

\section{Summary and discussions}
\label{225058_15Sep23}

\subsection{Summary}
In this work, we investigated the one-dimensional harmonic chain far
from equilibrium. We considered the four types of driving forces that do
not satisfy the detailed balance: (i) temporally correlated noise with
power-law spectrum $D(\omega)\sim \omega^{-2\theta}$, (ii)
conserving noise, (iii) periodic driving force, and (iv)
periodic deformation. For the driving force (i) with $\theta>-1/4$, the
model undergoes the anomalous diffusion ${\rm MSD}(t)\sim
t^{1/2+2\theta}$. On the contrary, for the driving forces (i) with
$\theta<-1/4$, and (ii)--(iv), {\rm MSD}(t) remains finite. As a
consequence, the crystalline order parameter has a finite value, unlike
the equilibrium systems where the Mermin-Wagner theorem prohibits the
long-range crystalline order in one dimension. 
We also discussed hyperuniformity of the density fluctuations in the
crystal phase.  We hope our work will stimulate further interest and
progress of the long-range
order~\cite{vicsek1995,toner1995,xy1995,reichl2010,nakano2021,loos2022long,leonardo2023,ikeda2023out}
and hyperuniformity~\cite{kim2018,oǧuz2019hyperuniformity,Duyu2022} in
non-equilibrium low-dimensional systems.

\subsection{Hyperuniformity}
For the driving forces (i) with $\theta<0$ and (iii), the power-spectrum
of the noise vanishes in the limit of the small frequency:
$\lim_{\omega\to 0}D_q(\omega)=0$.  This means that the
fluctuations of the noise are highly suppressed in a long time scale,
{\it i.e.}, the noise is temporally hyperuniform. For the driving force
(ii), $\lim_{q\to 0}D_q(\omega)=0$, implying that the noise is spatially
hyperuniform. For the driving force (iv),
$D_q(\omega)$ vanishes in the limits
$\omega\to 0$ and/or $q\to 0$, {\it i.e.}, the noise is
spatio-temporally hyperuniform. Our work demonstrated that these spatial
and temporal hyperuniformity of the noise yield hyperuniformity of the
density fluctuations~\cite{ikeda2023out}.

For more general and quantitative discussions, we consider the
spatio-temporally correlated noise whose Fourier spectrum is given by
for $\omega\ll 1$ and
$q\ll 1$
\begin{align}
D_q(\omega)\sim \omega^{-2\theta}q^{-2\rho}.\label{005925_6Nov23}
\end{align}
The driving force (i) corresponds to $\rho=0$, (ii)
corresponds to $\rho=-1$ and $\theta=0$, (iii) corresponds to $\rho=0$
and $\theta\to-\infty$, and (iv) corresponds to $\rho=-1$ and
$\theta\to-\infty$.  For the noise to be hyperuniform $\lim_{q\to
0,\omega\to 0}D_q(\omega)=0$, $\rho$ and $\theta$ should satisfy
$\rho\leq 0$, $\theta\leq 0$ and $(\rho,\theta)\neq (0,0)$.  The static
structure factor $S(q)$ for $q\ll 1$ is calculated
as~\cite{ikeda2023out}
\begin{align}
S(q)\approx q^2\ave{\tu_{q} \tu_{-q}} = 
\frac{q^2}{\pi}\int_0^{\infty}d\omega \frac{D_q(\omega)}{\omega^2+\lambda_q^2}
\sim
\begin{cases}
q^{-2\rho-4\theta} & \theta > -1/2 \\
q^{2-2\rho} &\theta\leq -1/2,
\end{cases}
 \label{110326_8Oct23}
\end{align}
where the phenomenological cut-off $\omega_c$ is needed to converge the
integral for $\theta\leq -1/2$, see Eq.~(\ref{134951_6Sep23}). The above
equation implies $\lim_{q\to 0}S(q)=0$ if the noise is hyperuniform.

\subsection{Anomalous diffusion of spatio-temporally correlated noise}
Here we briefly discuss the anomalous diffusion of a one-dimensional
system driven by the spatio-temporally correlated noise
Eq.~(\ref{005925_6Nov23}). For that purpose, we investigate the model in
the continuum limit Eq.~(\ref{020532_27Sep23}).  The scaling
transformations of Eq.~(\ref{020532_27Sep23}), $x\to bx$, $t\to
b^{z_t}$, $u\to b^{z_u}u$, lead to $z_t=2$ and $z_u = 1/2 + 2\theta +
\rho$~\cite{ikeda2023out}. Then, we get the anomalous diffusion ${\rm
MSD}\sim t^{2z_u/z_t}\sim t^{1/2+2\theta+\rho}$ for
$1/2+2\theta+\rho>0$.  For $1/2+2\theta+\rho<0$, on the contrary, the
diffusion is completely suppressed and the model has the long-range
crystalline order.

\subsection{Hyperuniformity and crystalline order}
For the existence of the crystalline order, $\ave{u_1^2}=\sum_{q}
\ave{\tu_{q}\tu_{-q}}/N$ should remain finite in the thermodynamic limit
$N\to\infty$. A necessary condition is $\ave{\tu_{q} \tu_{-q}}\propto
q^{-\mu}$ with $\mu<1$ for $q\ll 1$, which is tantamount to $S(q)\approx
q^2 \ave{\tu_{q}\tu_{-q}}\sim q^{2-\mu}$ for $q\ll 1$. In other words,
for the existence of the crystalline order in one dimension, the density
fluctuations should exhibit sufficiently strong hyperuniformity
$S(q)\sim q^\nu$ with $\nu>1$. This condition is more stringent than in
two-dimensional systems, where $\nu>0$ is enough to stabilize the
long-range crystalline order~\cite{leonardo2023}.

The generalization of the above argument to higher dimension $d$ is
straightforward. Let $\tilde{\bm{u}}(\bq)=\{\tu_a(\bq)\}_{a=1,\cdots,d}$
be the Fourier component of the displacement vector. Assuming that the
system is isotropic $\ave{\tu_a\tu_b}=\delta_{ab}\ave{\tu^2}$, one
obtains $S(\bq)\approx \abs{\bq}^2 \ave{\tu(\bq)\tu(-\bq)}$ for the harmonic
lattice in $d$ dimension, see Ref~\cite{kim2018}. Then, hyperuniformity
$S(\bq)\sim \abs{\bq}^{\nu}$ ($\nu>0$) implies $\ave{\tu(\bq)\tu(-\bq)}\approx 
\abs{\bq}^{-2}S(\bq)\sim \abs{\bq}^{\nu-2}$. To exist the long-range
crystalline order, the particles should localize around their lattice
positions. In other words, the mean-squared displacement from the
lattice position $\ave{u(\bx)^2}$ should remain finite. A rough
estimation of this quantity in $d$ dimension is~\cite{leonardo2023}
\begin{align}
\ave{u(\bx)^2} =\frac{1}{(2\pi)^d}\int d\bq \ave{\tu(\bq)\tu(-\bq)}
 \sim \int_0^{q_D} dq q^{d-3+\nu},\label{152459_3Oct23}
\end{align}
where $q_D$ denotes the Debye cut-off. Eq.~(\ref{152459_3Oct23}) remains
finite below the lower critical dimension 
\begin{align}
d_{\rm low}=2-\nu.\label{212906_9Oct23}
\end{align}
Therefore, the crystalline order can exist for $\nu>0$ in
$d=2$~\cite{leonardo2023}, and $\nu>1$ in $d=1$. The above argument also
implies that giant number fluctuations, $S(\bq)\sim \abs{\bq}^{\nu}$
with $\nu<0$, increases $d_{\rm low}$. Using Eq.~(\ref{110326_8Oct23}),
we get the lower critical dimension of the crystallization of the
systems driven by the spatio-temporally correlated noise
\begin{align}
d_{\rm low}=
\begin{cases}
 2+2\rho+4\theta & \theta > -1/2\\
 2\rho & \theta \leq -1/2
\end{cases}.
 \label{212237_9Oct23} 
\end{align}

\subsection{Comparison with $O(n)$ model}
In a previous work, we have investigated the $O(n)$ model driven by the
correlated noise of the noise spectrum $D(\omega,q)\sim
\omega^{-2\theta}q^{-2\rho}$~\cite{ikeda2023out}. For the model-A
dynamics~\cite{hohenberg1977}, the lower critical dimension for the
continuous symmetry breaking is $d_{\rm low}=2+2\rho+4\theta$ for
$\theta>-1/2$ and $d_{\rm low}=2\rho$ for $\theta\leq -1/2$, which
agrees with Eq.~(\ref{212237_9Oct23}). This is a reasonable result
because the order parameter of the crystallization is a non-conservative
quantity. On the contrary, since the density is a conservative
quantity~\cite{hohenberg1977}, the prediction for hyperuniformity
Eq.~(\ref{110326_8Oct23}) agrees with that of the model-B dynamics of
the $O(n)$ model. As a consequence, the relation between hyperuniformity
and $d_{\rm low}$, Eq.~(\ref{212906_9Oct23}), is not consistent with
ether the model-A and model-B dynamics of the $O(n)$
model~\cite{ikeda2023out}. This result highlights an essential
difference between the crystallization of particle systems and
ferromagnetic phase transition of lattice spin systems. Further studies
would be beneficial to elucidate the similarities and differences of
these models.

\subsection{Does fluid-solid transition occur in one dimension?}

Several non-equilibrium systems are known to exhibit phase transition in
one dimension. However, to the best of our knowledge, there are still
no known examples of continuous symmetry breaking in one dimension. This
work provides several promising candidates.

Our analysis for the driving forces (i) for $\theta<-1/4$ and (ii)-(iv)
showed that the crystalline order can emerge even in one dimension,
which is prohibited in equilibrium by the Mermin-Wagner theorem. One
natural question is whether the systems driven by these driving forces
exhibit liquid-solid phase transitions on increasing density. For the
harmonic potential studied in this manuscript, the dynamics of the
relative displacement $u_j$, Eq.~(\ref{dyn}), does not depend on the
lattice spacing $a$. Therefore, the qualitative behavior of the model is
also density-independent. What will happen for more realistic
interaction potentials, such as the Lennard-Jones
potential~\cite{hansen2013theory}, one-sided harmonic potential,
Hertzian potential~\cite{ohern2003}, and so on? Extensive numerical
simulations of systems driven by conserving
noise~\cite{hexner2017noise,leonardo2023}, chiral active
particles~\cite{Callegari2019,liebchen2022chiral}, and periodically
deforming particles~\cite{tjhung2017} for a wide range of density would
be beneficial to elucidate this point.

\section*{Acknowledgements}
We thank Y.~Nishikawa and Y. Kuroda for useful comments.


\paragraph{Funding information}
This project has received JSPS KAKENHI Grant Numbers 23K13031.

\bibliography{reference.bib}

\begin{thebibliography}{10}
\providecommand{\url}[1]{\texttt{#1}}
\providecommand{\urlprefix}{URL }
\expandafter\ifx\csname urlstyle\endcsname\relax
  \providecommand{\doi}[1]{doi:\discretionary{}{}{}#1}\else
  \providecommand{\doi}{doi:\discretionary{}{}{}\begingroup
  \urlstyle{rm}\Url}\fi
\providecommand{\eprint}[2][]{\url{#2}}

\bibitem{alexander1978}
S.~Alexander and P.~Pincus,
\newblock \emph{Diffusion of labeled particles on one-dimensional chains},
\newblock Phys. Rev. B \textbf{18}, 2011 (1978),
\newblock \doi{10.1103/PhysRevB.18.2011}.

\bibitem{karger1992}
J.~K\"arger,
\newblock \emph{Straightforward derivation of the long-time limit of the
  mean-square displacement in one-dimensional diffusion},
\newblock Phys. Rev. A \textbf{45}, 4173 (1992),
\newblock \doi{10.1103/PhysRevA.45.4173}.

\bibitem{kollmann2003}
M.~Kollmann,
\newblock \emph{Single-file diffusion of atomic and colloidal systems:
  Asymptotic laws},
\newblock Phys. Rev. Lett. \textbf{90}, 180602 (2003),
\newblock \doi{10.1103/PhysRevLett.90.180602}.

\bibitem{lin2005}
B.~Lin, M.~Meron, B.~Cui, S.~A. Rice and H.~Diamant,
\newblock \emph{From random walk to single-file diffusion},
\newblock Phys. Rev. Lett. \textbf{94}, 216001 (2005),
\newblock \doi{10.1103/PhysRevLett.94.216001}.

\bibitem{taloni2017single}
A.~Taloni, O.~Flomenbom, R.~Casta{\~{n}}eda-Priego and F.~Marchesoni,
\newblock \emph{Single file dynamics in soft materials},
\newblock Soft Matter \textbf{13}(6), 1096 (2017),
\newblock \doi{10.1039/c6sm02570f}.

\bibitem{tirthankar2022}
T.~Banerjee, R.~L. Jack and M.~E. Cates,
\newblock \emph{Role of initial conditions in one-dimensional diffusive
  systems: Compressibility, hyperuniformity, and long-term memory},
\newblock Phys. Rev. E \textbf{106}, L062101 (2022),
\newblock \doi{10.1103/PhysRevE.106.L062101}.

\bibitem{zwanzig2001nonequilibrium}
R.~Zwanzig,
\newblock \emph{Nonequilibrium statistical mechanics},
\newblock Oxford university press (2001).

\bibitem{ashcroft2022solid}
N.~W. Ashcroft and N.~D. Mermin,
\newblock \emph{Solid state physics},
\newblock Cengage Learning (2022).

\bibitem{altland2010condensed}
A.~Altland and B.~D. Simons,
\newblock \emph{Condensed matter field theory},
\newblock Cambridge university press (2010).

\bibitem{mermin1968}
N.~D. Mermin,
\newblock \emph{Crystalline order in two dimensions},
\newblock Phys. Rev. \textbf{176}, 250 (1968),
\newblock \doi{10.1103/PhysRev.176.250}.

\bibitem{mermin1969}
N.~D. Mermin and H.~Wagner,
\newblock \emph{Absence of ferromagnetism or antiferromagnetism in one- or
  two-dimensional isotropic heisenberg models},
\newblock Phys. Rev. Lett. \textbf{17}, 1133 (1966),
\newblock \doi{10.1103/PhysRevLett.17.1133}.

\bibitem{torquato2018hyperuniform}
S.~Torquato,
\newblock \emph{Hyperuniform states of matter},
\newblock Physics Reports \textbf{745}, 1 (2018),
\newblock \doi{10.1016/j.physrep.2018.03.001}.

\bibitem{kim2018}
J.~Kim and S.~Torquato,
\newblock \emph{Effect of imperfections on the hyperuniformity of many-body
  systems},
\newblock Phys. Rev. B \textbf{97}, 054105 (2018),
\newblock \doi{10.1103/PhysRevB.97.054105}.

\bibitem{oguz2017}
E.~C. O\ifmmode~\breve{g}\else \u{g}\fi{}uz, J.~E.~S. Socolar, P.~J. Steinhardt
  and S.~Torquato,
\newblock \emph{Hyperuniformity of quasicrystals},
\newblock Phys. Rev. B \textbf{95}, 054119 (2017),
\newblock \doi{10.1103/PhysRevB.95.054119}.

\bibitem{o^^c7^^a7uz2019hyperuniformity}
E.~O^^c7^^a7uz, J.~E. Socolar, P.~J. Steinhardt and S.~Torquato,
\newblock \emph{Hyperuniformity and anti-hyperuniformity in one-dimensional
  substitution tilings},
\newblock Acta Crystallographica Section A: Foundations and Advances
  \textbf{75}(1), 3 (2019),
\newblock \doi{10.1107/S2053273318015528}.

\bibitem{feynman1954}
R.~P. Feynman,
\newblock \emph{Atomic theory of the two-fluid model of liquid helium},
\newblock Phys. Rev. \textbf{94}, 262 (1954),
\newblock \doi{10.1103/PhysRev.94.262}.

\bibitem{reatto1967}
L.~Reatto and G.~V. Chester,
\newblock \emph{Phonons and the properties of a bose system},
\newblock Phys. Rev. \textbf{155}, 88 (1967),
\newblock \doi{10.1103/PhysRev.155.88}.

\bibitem{Torquato2008}
S.~Torquato, A.~Scardicchio and C.~E. Zachary,
\newblock \emph{Point processes in arbitrary dimension from fermionic gases,
  random matrix theory, and number theory},
\newblock Journal of Statistical Mechanics: Theory and Experiment
  \textbf{2008}(11), P11019 (2008),
\newblock \doi{10.1088/1742-5468/2008/11/p11019}.

\bibitem{weijs2015emergent}
J.~H. Weijs, R.~Jeanneret, R.~Dreyfus and D.~Bartolo,
\newblock \emph{Emergent hyperuniformity in periodically driven emulsions},
\newblock Phys. Rev. Lett. \textbf{115}, 108301 (2015),
\newblock \doi{10.1103/PhysRevLett.115.108301}.

\bibitem{lei2019hydrodynamics}
Q.-L. Lei and R.~Ni,
\newblock \emph{Hydrodynamics of random-organizing hyperuniform fluids},
\newblock Proceedings of the National Academy of Sciences \textbf{116}(46),
  22983 (2019),
\newblock \doi{10.1073/pnas.1911596116}.

\bibitem{huang2021circular}
M.~Huang, W.~Hu, S.~Yang, Q.-X. Liu and H.~P. Zhang,
\newblock \emph{Circular swimming motility and disordered hyperuniform state in
  an algae system},
\newblock Proceedings of the National Academy of Sciences \textbf{118}(18)
  (2021),
\newblock \doi{10.1073/pnas.2100493118}.

\bibitem{hyperchiral2022}
B.~Zhang and A.~Snezhko,
\newblock \emph{Hyperuniform active chiral fluids with tunable internal
  structure},
\newblock Phys. Rev. Lett. \textbf{128}, 218002 (2022),
\newblock \doi{10.1103/PhysRevLett.128.218002}.

\bibitem{kuroda2023microscopic}
Y.~Kuroda and K.~Miyazaki,
\newblock \emph{Microscopic theory for hyperuniformity in two-dimensional
  chiral active fluid},
\newblock Journal of Statistical Mechanics: Theory and Experiment
  \textbf{2023}(10), 103203 (2023),
\newblock \doi{10.1088/1742-5468/ad0639}.

\bibitem{hexner2015}
D.~Hexner and D.~Levine,
\newblock \emph{Hyperuniformity of critical absorbing states},
\newblock Phys. Rev. Lett. \textbf{114}, 110602 (2015),
\newblock \doi{10.1103/PhysRevLett.114.110602}.

\bibitem{hexner2017noise}
D.~Hexner and D.~Levine,
\newblock \emph{Noise, diffusion, and hyperuniformity},
\newblock Phys. Rev. Lett. \textbf{118}, 020601 (2017),
\newblock \doi{10.1103/PhysRevLett.118.020601}.

\bibitem{leonardo2023}
L.~Galliano, M.~E. Cates and L.~Berthier,
\newblock \emph{Two-dimensional crystals far from equilibrium},
\newblock Phys. Rev. Lett. \textbf{131}, 047101 (2023),
\newblock \doi{10.1103/PhysRevLett.131.047101}.

\bibitem{Lei2019}
Q.-L. Lei, M.~P. Ciamarra and R.~Ni,
\newblock \emph{Nonequilibrium strongly hyperuniform fluids of circle active
  particles with large local density fluctuations},
\newblock Science Advances \textbf{5}(1) (2019),
\newblock \doi{10.1126/sciadv.aau7423}.

\bibitem{ikeda2023out}
H.~Ikeda,
\newblock \emph{Correlated noise and critical dimensions},
\newblock Phys. Rev. E \textbf{108}, 064119 (2023),
\newblock \doi{10.1103/PhysRevE.108.064119}.

\bibitem{Callegari2019}
A.~Callegari and G.~Volpe,
\newblock \emph{Numerical Simulations of Active Brownian Particles}, pp.
  211--238,
\newblock Springer International Publishing, Cham,
\newblock ISBN 978-3-030-23370-9,
\newblock \doi{10.1007/978-3-030-23370-9_7} (2019).

\bibitem{liebchen2022chiral}
B.~Liebchen and D.~Levis,
\newblock \emph{Chiral active matter},
\newblock Europhysics Letters \textbf{139}(6), 67001 (2022),
\newblock \doi{10.1209/0295-5075/ac8f69}.

\bibitem{tjhung2017}
E.~Tjhung and L.~Berthier,
\newblock \emph{Discontinuous fluidization transition in time-correlated
  assemblies of actively deforming particles},
\newblock Phys. Rev. E \textbf{96}, 050601 (2017),
\newblock \doi{10.1103/PhysRevE.96.050601}.

\bibitem{ikeda2023does}
H.~Ikeda and Y.~Kuroda,
\newblock \emph{Does spontaneous symmetry breaking occur in periodically driven
  low-dimensional non-equilibrium classical systems?},
\newblock arXiv preprint arXiv:2304.14235  (2023),
\newblock \doi{10.48550/arXiv.2304.14235}.

\bibitem{nishimori2010elements}
H.~Nishimori and G.~Ortiz,
\newblock \emph{Elements of phase transitions and critical phenomena},
\newblock Oup Oxford (2010).

\bibitem{greiner2012thermodynamics}
W.~Greiner, L.~Neise and H.~St{\"o}cker,
\newblock \emph{Thermodynamics and statistical mechanics},
\newblock Springer Science \& Business Media (2012).

\bibitem{dolai2020universal}
P.~Dolai, A.~Das, A.~Kundu, C.~Dasgupta, A.~Dhar and K.~V. Kumar,
\newblock \emph{Universal scaling in active single-file dynamics},
\newblock Soft Matter \textbf{16}(30), 7077 (2020),
\newblock \doi{10.1039/d0sm00687d}.

\bibitem{per1987}
P.~Bak, C.~Tang and K.~Wiesenfeld,
\newblock \emph{Self-organized criticality: An explanation of the 1/f noise},
\newblock Phys. Rev. Lett. \textbf{59}, 381 (1987),
\newblock \doi{10.1103/PhysRevLett.59.381}.

\bibitem{Newman2005}
M.~Newman,
\newblock \emph{Power laws, pareto distributions and zipf{\textquotesingle}s
  law},
\newblock Contemporary Physics \textbf{46}(5), 323 (2005),
\newblock \doi{10.1080/00107510500052444}.

\bibitem{milotti2002}
E.~Milotti,
\newblock \emph{1/f noise: a pedagogical review},
\newblock arXiv preprint physics/0204033  (2002),
\newblock \doi{10.48550/arXiv.physics/0204033}.

\bibitem{eliazar2009unified}
I.~Eliazar and J.~Klafter,
\newblock \emph{A unified and universal explanation for l{\'{e}}vy laws and 1/f
  noises},
\newblock Proceedings of the National Academy of Sciences \textbf{106}(30),
  12251 (2009),
\newblock \doi{10.1073/pnas.0900299106}.

\bibitem{burger1989}
E.~Medina, T.~Hwa, M.~Kardar and Y.-C. Zhang,
\newblock \emph{Burgers equation with correlated noise: Renormalization-group
  analysis and applications to directed polymers and interface growth},
\newblock Phys. Rev. A \textbf{39}, 3053 (1989),
\newblock \doi{10.1103/PhysRevA.39.3053}.

\bibitem{chen2021stone}
D.~Chen, Y.~Zheng, L.~Liu, G.~Zhang, M.~Chen, Y.~Jiao and H.~Zhuang,
\newblock \emph{Stone{\textendash}wales defects preserve hyperuniformity in
  amorphous two-dimensional networks},
\newblock Proceedings of the National Academy of Sciences \textbf{118}(3)
  (2021),
\newblock \doi{10.1073/pnas.2016862118}.

\bibitem{imry1975}
Y.~Imry and S.-k. Ma,
\newblock \emph{Random-field instability of the ordered state of continuous
  symmetry},
\newblock Phys. Rev. Lett. \textbf{35}, 1399 (1975),
\newblock \doi{10.1103/PhysRevLett.35.1399}.

\bibitem{narayan2007long}
V.~Narayan, S.~Ramaswamy and N.~Menon,
\newblock \emph{Long-lived giant number fluctuations in a swarming granular
  nematic},
\newblock Science \textbf{317}(5834), 105 (2007),
\newblock \doi{10.1126/science.1140414}.

\bibitem{Zhang2010}
H.~P. Zhang, A.~Be'er, E.-L. Florin and H.~L. Swinney,
\newblock \emph{Collective motion and density fluctuations in bacterial
  colonies},
\newblock Proceedings of the National Academy of Sciences \textbf{107}(31),
  13626 (2010),
\newblock \doi{10.1073/pnas.1001651107}.

\bibitem{ro2021}
S.~Ro, Y.~Kafri, M.~Kardar and J.~Tailleur,
\newblock \emph{Disorder-induced long-ranged correlations in scalar active
  matter},
\newblock Phys. Rev. Lett. \textbf{126}, 048003 (2021),
\newblock \doi{10.1103/PhysRevLett.126.048003}.

\bibitem{Zehnder2015}
S.~M. Zehnder, M.~Suaris, M.~M. Bellaire and T.~E. Angelini,
\newblock \emph{Cell volume fluctuations in {MDCK} monolayers},
\newblock Biophysical Journal \textbf{108}(2), 247 (2015),
\newblock \doi{10.1016/j.bpj.2014.11.1856}.

\bibitem{ohern2003}
C.~S. O'Hern, L.~E. Silbert, A.~J. Liu and S.~R. Nagel,
\newblock \emph{Jamming at zero temperature and zero applied stress: The
  epitome of disorder},
\newblock Phys. Rev. E \textbf{68}, 011306 (2003),
\newblock \doi{10.1103/PhysRevE.68.011306}.

\bibitem{goldstein2002classical}
H.~Goldstein, C.~Poole and J.~Safko,
\newblock \emph{Classical mechanics} (2002).

\bibitem{Tjhung2017excitation}
E.~Tjhung and T.~Kawasaki,
\newblock \emph{Excitation of vibrational soft modes in disordered systems
  using active oscillation},
\newblock Soft Matter \textbf{13}(1), 111 (2017),
\newblock \doi{10.1039/c6sm00788k}.

\bibitem{aikeda2015}
A.~Ikeda and L.~Berthier,
\newblock \emph{Thermal fluctuations, mechanical response, and hyperuniformity
  in jammed solids},
\newblock Phys. Rev. E \textbf{92}, 012309 (2015),
\newblock \doi{10.1103/PhysRevE.92.012309}.

\bibitem{vicsek1995}
T.~Vicsek, A.~Czir\'ok, E.~Ben-Jacob, I.~Cohen and O.~Shochet,
\newblock \emph{Novel type of phase transition in a system of self-driven
  particles},
\newblock Phys. Rev. Lett. \textbf{75}, 1226 (1995),
\newblock \doi{10.1103/PhysRevLett.75.1226}.

\bibitem{toner1995}
J.~Toner and Y.~Tu,
\newblock \emph{Long-range order in a two-dimensional dynamical $\mathrm{XY}$
  model: How birds fly together},
\newblock Phys. Rev. Lett. \textbf{75}, 4326 (1995),
\newblock \doi{10.1103/PhysRevLett.75.4326}.

\bibitem{xy1995}
K.~E. Bassler and Z.~R\'acz,
\newblock \emph{Existence of long-range order in the steady state of a
  two-dimensional, two-temperature xy model},
\newblock Phys. Rev. E \textbf{52}, R9 (1995),
\newblock \doi{10.1103/PhysRevE.52.R9}.

\bibitem{reichl2010}
M.~D. Reichl, C.~I. Del~Genio and K.~E. Bassler,
\newblock \emph{Phase diagram for a two-dimensional, two-temperature, diffusive
  $xy$ model},
\newblock Phys. Rev. E \textbf{82}, 040102 (2010),
\newblock \doi{10.1103/PhysRevE.82.040102}.

\bibitem{nakano2021}
H.~Nakano, Y.~Minami and S.-i. Sasa,
\newblock \emph{Long-range phase order in two dimensions under shear flow},
\newblock Phys. Rev. Lett. \textbf{126}, 160604 (2021),
\newblock \doi{10.1103/PhysRevLett.126.160604}.

\bibitem{loos2022long}
S.~A.~M. Loos, S.~H.~L. Klapp and T.~Martynec,
\newblock \emph{Long-range order and directional defect propagation in the
  nonreciprocal $\mathit{XY}$ model with vision cone interactions},
\newblock Phys. Rev. Lett. \textbf{130}, 198301 (2023),
\newblock \doi{10.1103/PhysRevLett.130.198301}.

\bibitem{Duyu2022}
D.~Chen, Y.~Liu, Y.~Zheng, H.~Zhuang, M.~Chen and Y.~Jiao,
\newblock \emph{Disordered hyperuniform quasi-one-dimensional materials},
\newblock Phys. Rev. B \textbf{106}, 235427 (2022),
\newblock \doi{10.1103/PhysRevB.106.235427}.

\bibitem{hohenberg1977}
P.~C. Hohenberg and B.~I. Halperin,
\newblock \emph{Theory of dynamic critical phenomena},
\newblock Rev. Mod. Phys. \textbf{49}, 435 (1977),
\newblock \doi{10.1103/RevModPhys.49.435}.

\bibitem{hansen2013theory}
J.-P. Hansen and I.~R. McDonald,
\newblock \emph{Theory of simple liquids: with applications to soft matter},
\newblock Academic press (2013).

\end{thebibliography}

\nolinenumbers

\end{document}